\pdfoutput=1

\documentclass[11pt]{article}

\usepackage[final]{acl}

\usepackage{times}
\usepackage{latexsym}
\usepackage{adjustbox}
\usepackage{hyperref}       
\usepackage{url}            
\usepackage{booktabs}       
\usepackage{amsfonts}       
\usepackage{nicefrac}       
\usepackage{xcolor}         
\usepackage{graphicx}
\usepackage{bbding}
\usepackage{multirow}
\usepackage{wrapfig}
\usepackage{amsmath}
\usepackage{amssymb}
\usepackage{mathtools}
\usepackage{amsthm}
\usepackage{array}
\usepackage{arydshln}
\usepackage{float}
\usepackage{pifont}
\newcommand{\cmark}{\ding{51}}%
\newcommand{\xmark}{\ding{55}}%
\usepackage{hyperref}
\usepackage{url}
\usepackage{booktabs}
\usepackage{algorithm}
\usepackage{algorithmic}
\usepackage{bbding}
\usepackage{amsfonts} 
\usepackage{mathtools}
\usepackage{amsthm}
\usepackage{array}
\usepackage{subfig} 
\usepackage{enumitem}
\usepackage{arydshln}
\usepackage{microtype}
\usepackage{epsfig}
\usepackage{bm}
\usepackage{bbding}
\usepackage{pifont}

\usepackage{dsfont}
\usepackage{color}
\usepackage{tabularx}

\usepackage[T1]{fontenc}

\usepackage[utf8]{inputenc}

\usepackage{microtype}

\usepackage{inconsolata}

\usepackage{graphicx}

\title{SongComposer: A Large Language Model for Lyric and Melody Generation in Song Composition}

\author{
 \textbf{Shuangrui Ding\textsuperscript{1}}\thanks{equal contribution},
 \textbf{Zihan Liu\textsuperscript{2,3}}$^*$,
 \textbf{Xiaoyi Dong\textsuperscript{1,3}},
 \textbf{Pan Zhang\textsuperscript{3}},
\\
 \textbf{Rui Qian\textsuperscript{1}},
 \textbf{Junhao Huang\textsuperscript{3}},
 \textbf{Conghui He\textsuperscript{3}},
 \textbf{Dahua Lin\textsuperscript{1,3,4}},
 \textbf{Jiaqi Wang\textsuperscript{3}}
\\
 \textsuperscript{1}The Chinese University of Hong Kong,
 \textsuperscript{2}Beihang University, \\
 \textsuperscript{3}Shanghai AI Laboratory, 
 \textsuperscript{4}CPII under InnoHK
\\
 \small{
   \textbf{Correspondence:} \href{mailto:wangjiaqi@pjlab.org.cn}{wangjiaqi@pjlab.org.cn}
 }
}

\begin{document}
\maketitle

\begin{figure*}[h]
    \centering
    \includegraphics[width=\textwidth]{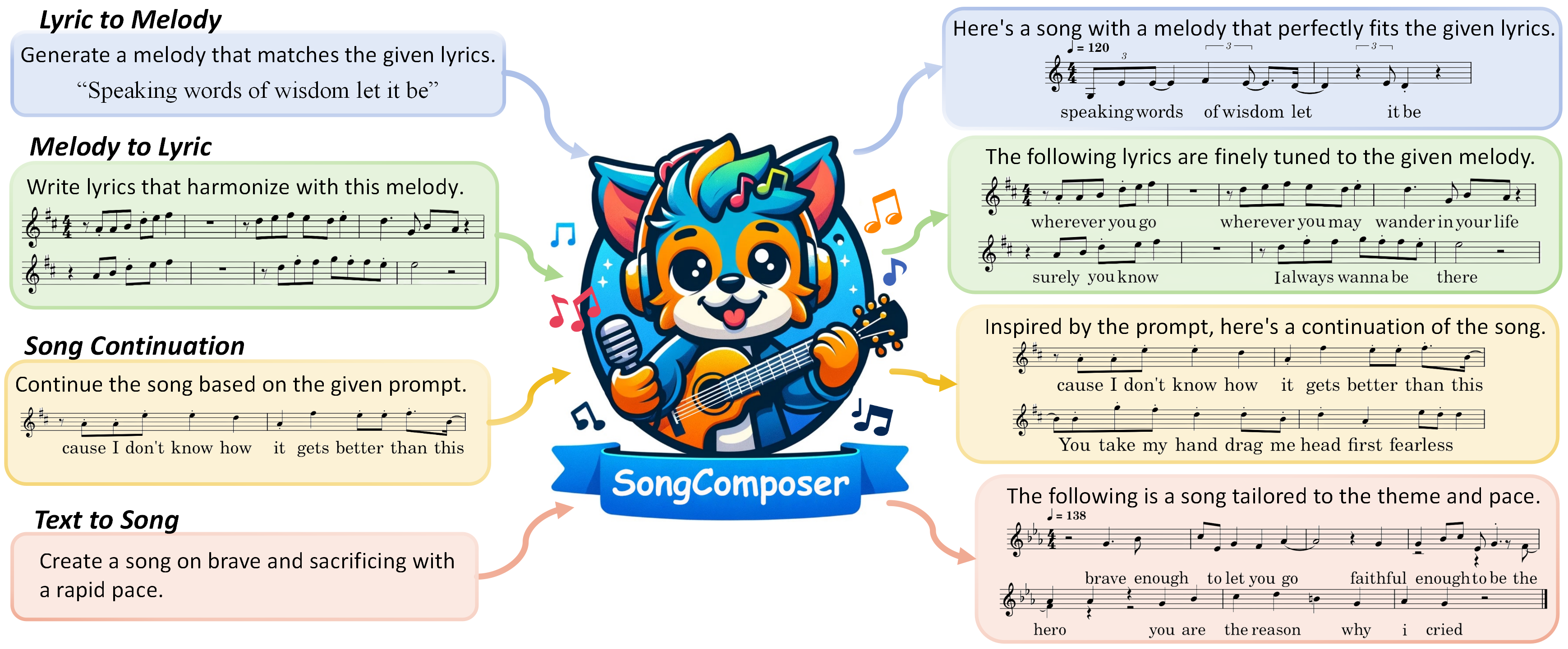}
    \captionof{figure}{Overview of the song-related instruction-following composition by SongComposer. SongComposer utilizes symbolic song representation to compose melodies tailored to lyrics, craft lyrics to complement melodies, extend existing songs, and generate new songs from textual prompts.}
    \label{fig:framework}
    \vspace{-15pt}
\end{figure*}

\begin{abstract}

Creating lyrics and melodies for the vocal track in a symbolic format, known as song composition, demands expert musical knowledge of melody, an advanced understanding of lyrics, and precise alignment between them. Despite achievements in sub-tasks such as lyric generation, lyric-to-melody, and melody-to-lyric, etc, a unified model for song composition has not yet been achieved. In this paper, we introduce SongComposer, \textbf{a pioneering step towards a unified song composition model that can readily create symbolic lyrics and melodies following instructions.} SongComposer is a music-specialized large language model (LLM) that, for the first time, integrates the capability of simultaneously composing lyrics and melodies into LLMs by leveraging three key innovations:
1) a flexible tuple format for word-level alignment of lyrics and melodies, 2) an extended tokenizer vocabulary for song notes, with scalar initialization based on musical knowledge to capture rhythm, and 3) a multi-stage pipeline that captures musical structure, starting with motif-level melody patterns and progressing to phrase-level structure for improved coherence. 
Extensive experiments demonstrate that SongComposer outperforms advanced LLMs, including GPT-4, in tasks such as lyric-to-melody generation, melody-to-lyric generation, song continuation, and text-to-song creation.
We showcase the generated samples on our project page\footnote{\url{https://pjlab-songcomposer.github.io/}}.
Moreover, we will release SongCompose, a large-scale dataset for training, containing paired lyrics and melodies in Chinese and English.

\end{abstract}

\section{Introduction}

Symbolic song composition aims to generate the vocal track of a song as a sequence of symbols representing lyrics and melodies. It is a vital task in song generation and requires professional knowledge.
Recently, this field has become a highly active area of research in both academic and industrial domains. 
Previous efforts have made significant progress in isolated sub-tasks of song composition such as lyric generation~\citep{zhang2022youling}, lyric-to-melody~\citep{ConditionalL2M,ju2022telemelody,sheng2021songmass, ReLyMeL2M} or melody-to-lyric generation~\citep{sheng2021songmass, ma202lyricist}.
However, the absence of a unified framework for generating both lyrics and melodies concurrently while adhering to specific instructions poses a challenge for seamless adaptation, thereby creating a higher hurdle for everyday amateurs.

The recent surge in large language models (LLMs) has dramatically revolutionized the artificial intelligence landscape, especially in natural language understanding and generation~\citep{brown2020language, vicuna2023, wei2021finetuned, chowdhery2023palm, raffel2020exploring, devlin2018bert}. These models have established new benchmarks for parsing and producing human language, showcasing human-level proficiency in complex language environments. Given that symbolic song representation shares structural similarities with human language, it seems plausible that LLMs could facilitate the creation of symbolic songs. 
Furthermore, unlike previous non-LLM methods~\citep{sheng2021songmass, ju2022telemelody} that handle only specific tasks, LLMs can integrate various sub-tasks of song composition into a single model due to their instruction-following abilities.

\begin{figure*}
    \centering
    \includegraphics[width=\linewidth]{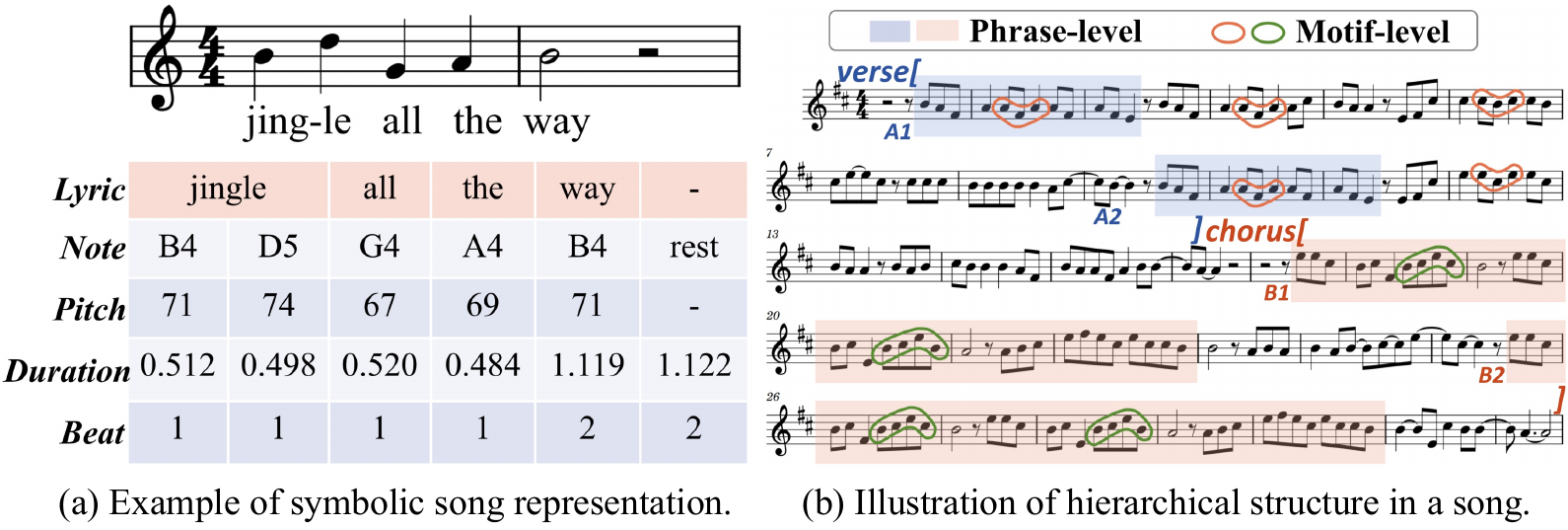}
    \caption{(a) Symbolic song representation involves precise alignment of notes and lyrics; (b) The structure of a song often comprises motif-level and phrase-level concepts.}
    \vspace{-15pt}
    \label{fig:teaser}
\end{figure*}

However, enabling LLMs to compose full-length songs that harmonize melody and lyrics is not a trivial task. 
First, as illustrated in Figure~\ref{fig:teaser}(a), symbolic song representation would decompose a song into its lyrics and note attributes (pitch, beat) and form a strict word-level alignment. Therefore, aligning lyric and melody attributes in a unified and efficient manner for LLMs is indispensable yet remains unexplored.
Secondly, a song typically features a well-organized and hierarchical structure~\citep{dai2022missing}. For example, a composer usually uses the concept of motif and phrase to enrich the unity of a song. A motif is a recurring musical idea that serves as a fundamental building unit, and a phrase is a broader segment of music that forms a complete thought or expression. As shown in Figure~\ref{fig:teaser}(b), a single song may have a clear high-level phrase structure like Verse-Chorus, and across the whole song, there may be repetitive patterns known as motifs. Thus, augmenting LLMs to understand these succinct musical structures is of vital importance and may require explicitly curated knowledge input and design. 
Last but not least, current symbolic song datasets~\citep{yu2021conditional, wang2022opencpop, huang2021multi} are either limited in quantity or lacking in quality. They often miss precise alignments between melody and lyrics, impeding progress in symbolic song generation.

To address the aforementioned challenges, we introduce SongComposer, an LLM capable of generating whole-song compositions that harmoniously integrate both melodies and lyrics. To the best of our knowledge, this is the first attempt to generate lyrics and melody simultaneously using LLMs.

Specifically, we propose a word-level tuple format to construct melody and lyric attributes in a flexible and unified manner, providing an efficient interface for aligning melody and lyrics.
Besides, we introduce a scalar initialization method to seamlessly initialize pitch tokens based on the existing vocabulary of LLMs. This method initializes a central pitch first and then sets the remaining note pitches as multiples of the central pitch embeddings. In this way, we explicitly introduce and reinforce the relationship between pitches to LLMs.

To learn the hierarchical structure of a song, we use a progressive training approach with SongComposer, enabling the model to recognize patterns of motifs and phrases. 
Initially, we extract highly repetitive melody snippets and treat these as general motifs for motif-level melody training. Subsequently, we insert special tokens to denote phrase concepts when training on the full-length song data, instructing the model to directly identify which parts of the song correspond to verses, choruses, or other phrases.
Based on these designs, our model is encouraged to generate structure-aware compositions that exhibit motif-level and phrase-level coherence.
 
Regarding the dataset, we have carefully compiled and curated a comprehensive high-quality dataset, SongCompose. This dataset comprises 280K songs with pure lyrics, 20K sets of pure melodies, and 8K paired lyrics and melodies in both Chinese and English. 
Moreover, it covers not only the pretraining dataset but also the supervised fine-tuning dataset for LLMs.
Notably, the paired data feature precise word-level alignment, and this portion has been curated from scratch. We believe this large-scale dataset can serve as a critical resource for training large language models, and we will release it to propel further research in this field.

We evaluate SongComposer on four song-related tasks, as shown in Figure~\ref{fig:framework}. Extensive experiments demonstrate that SongComposer outperforms advanced GPT-4 and several open-source LLMs both in terms of quality and adherence to the prompt. Moreover, we excel in the traditional model~\citep{sheng2021songmass, ju2022telemelody} on specific lyric-to-melody tasks.
In addition, we conduct a thorough ablation study to verify the effectiveness of the proposed components. We also include a memorization test \citep{carlini2022quantifying, agostinelli2023musiclm} to check for inappropriate copying from the dataset, revealing that SongComposer's output significantly differs from the original sequences in the pretraining dataset.

In short, our contributions are as follows:
\begin{itemize}[itemsep=3pt,topsep=0pt,parsep=0pt,leftmargin=16pt]
    \item We introduce SongComposer, an LLM capable of generating whole-song singable sheets that include both melodies and lyrics with well-structured formats following instructions.
    \item We propose a novel scalar initialization for note pitches and integrate motif- and phrase-level knowledge to enhance the model's understanding of pitch attributes and song structure.
    \item Extensive experiments show SongComposer outperforms traditional composition models and advanced LLMs like GPT-4 in various song-related generation tasks.
\end{itemize}

\section{Related Work}
\noindent\textbf{Symbolic Song Composition} involves key tasks such as generating song lyrics, composing melodies, and producing lyrics-melody pairs. Lyric generation aims to create meaningful and coherent lyrics using deep learning techniques~\cite{malmi2016dopelearning, zhang2022youling, xue2021deeprapper}. Melody generation~\cite{wu2019hierarchical, colombo2017deep} focuses on autonomously composing melodies that can stand alone. Lyric-to-melody generation~\cite{ConditionalL2M, ju2022telemelody, sheng2021songmass, ReLyMeL2M} takes it further by generating melodies that align with the given lyrics. The reverse task, melody-to-lyrics generation~\cite{bao2019neural, li2020rigid, sheng2021songmass}, involves creating lyrics that match a given melody. While these methods are effective for specific tasks, they typically cannot handle comprehensive composition with a single model. In contrast, SongComposer can simultaneously process both lyrics and melodies in a unified format, leveraging the power of LLMs.


\noindent\textbf{Large Language Models}~\cite{raffel2020exploring, radford2018improving, chowdhery2023palm, touvron2023llama, openai2023gpt4, ouyang2022training, openai2022chatgpt, ouyang2022training, vicuna2023, guo2025deepseek} have significantly enhanced natural language processing, showcasing impressive capabilities across diverse tasks. 
In the domain of symbolic music creation, recent endeavors~\cite{yuan2024chatmusician,deng2024composerx} propose employing large language models for generating symbolic pure music. However, crafting compositions encompassing both lyrics and melodies with LLMs remains an open problem. Inspired by the powerful human-level language capabilities of LLMs, we have developed the first unified LLMs framework that expands their application to lyric and melody composition for song generation.

\noindent\textbf{Paired Lyric-Melody Singing Dataset} is crucial for song generation. Existing datasets like JVS-MuSiC~\cite{tamaru2020jvs}, PopCS~\cite{liu2022diffsinger}, and OpenSinger~\cite{huang2021multi} offer diverse singing data but lack proper lyric-melody alignment. Datasets such as NUS-48E~\cite{duan2013nus}, NHSS~\cite{sharma2021nhss}, Tohoku Kiritan~\cite{ogawa2021tohoku}, and Opencpop~\cite{wang2022opencpop} provide aligned corpora in multiple languages but are limited in scale and style diversity. Recently, M4Singer~\cite{zhang2022m4singer} compiled around 700 Chinese songs, yet this is still insufficient for training an LLM for symbolic music generation. In this work, we collect approximately 8K symbolic songs in English and Chinese from scratch to train SongComposer.

\section{SongComposer}
\subsection{Symbolic Representation for LLMs}
\label{sec:format}

\noindent\textbf{Pure Melody Format.}
Inspired by the beat-based REMI representation \citep{huang2020pop}, we first decompose the notes into three symbolic attributes: note pitch \( p \), note duration \( d \), and rest duration \( r \). The pitch range \( p \) is from MIDI note numbers 48 to 83, corresponding to notes C3 to B5, which is the most common range for human vocal performance. 
Given the tempo of the melody, measured in beats per minute (bpm), we measure the note duration \( d \in \mathbb{Z} \) and rest duration \( r \in \mathbb{Z} \) in the number of 1/16 beat:
\begin{align*}
    d_k &= \phi(\frac{\text{bpm}}{60}(\text{note-end}_k - \text{note-start}_k) \times 16), \\
    r_k &= \phi(\frac{\text{bpm}}{60}(\text{note-start}_{k+1} - \text{note-end}_k) \times 16),
\end{align*}
where \text{note-start} and \text{note-end} are times in seconds, \( k \) denotes the note index number and $\phi(\cdot)$ is an operator that constrains the value to the nearest integer within the range \( [1, 256] \).

Then each note of pure melody is formatted in a tuple as follows:
\[
\small
\begin{array}{l}``\langle\text{bom}\rangle \text{ bpm is } \{bpm\}. \text{ Total } \{num\} \text{ lines.} \\
\text{The 1-st line:} \, \langle p_1 \rangle, d_1 \,|\, \langle \text{rest} \rangle, r_1 \,|\, \langle p_2 \rangle, d_2 \,|\, \langle \text{rest} \rangle, r_2 \cdots \\
\text{The 2-nd line:} \, \cdots \, \langle\text{eom}\rangle"
\end{array}
\]

\noindent where we treat $\langle \text{rest} \rangle$ as a type of note and skip the rest tuple if $r < 8$. $\langle \text{bom} \rangle$ and $\langle \text{eom} \rangle$ indicate the beginning and end of the melody, respectively. Note that $\langle \cdot \rangle$ represents special tokens we add outside the existing vocabulary. 

\noindent\textbf{Pure Lyric Format.}
The lyrics share the same language as LLMs, thus it can be directly used without additional design. The input of the pure lyric is formatted as follows:
\[
\small
\begin{array}{l}
``\langle\text{bol}\rangle \text{ Chinese/English song. Total } \{num\} \text{ lines.} \\
\text{The 1-st line:} \, w_1 \, w_2 \, \cdots \\
\text{The 2-nd line:} \, \cdots \, \langle\text{eol}\rangle"
\end{array}
\]

\noindent where special tokens $\langle\text{bol}\rangle$ and $\langle\text{eol}\rangle$ indicate the beginning and the end of pure lyrics, respectively; $w$ denotes a word in the lyrics.

\noindent\textbf{Paired Data Format.} 
We use word-level alignment to combine pure melody and lyrics as for paired data format. 
Formally, the input of the word-level paired melody is formatted as follows:
\[
\small
\begin{array}{l}
``\langle\text{bop}\rangle \text{ Chinese/English song.} \text{ bpm is } \{bpm\}. \\
\text{Total } \{num\} \text{ lines.} \\
\text{The 1-st line:} \, \langle p_1 \rangle, d_1, w_1 \,|\, \langle \text{rest} \rangle, r_1 \,|\, \langle p_2 \rangle, d_2, w_2 \,|\, \langle \text{rest} \rangle, r_2 \cdots \\
\text{The 2-nd line:} \,  \cdots \, \langle\text{eop}\rangle"
\end{array}
\]

\noindent where special tokens $\langle\text{bop}\rangle$ and $\langle\text{eop}\rangle$ indicate the beginning and the end of pair data. When a single lyric word is sung to multiple musical notes, we add a numerical suffix to the word to specify which note the word corresponds to. We show the examples of each proposed format in Appendix~\ref{app:tuple}.

\subsection{Pitch Initialization} 
\label{sec:pitch}
Motivated by the strong logical and mathematical relationship between different pitches, we argue that initializing pitch tokens with a strong prior on their relationships would be beneficial for the model to interpret pitch elements.
Therefore, we attempt four initialization methods for pitch tokens to verify our intuition. 

\noindent\textbf{Average Initialization} creates the embedding for new pitch tokens $\langle p \rangle$ by averaging the existing token embeddings of left bracket ($\langle$), pitch number ($p$), and right bracket ($\rangle$).

\noindent\textbf{Gaussian Initialization} generates embeddings for new pitch tokens using a Gaussian distribution, with the mean and variance calculated from existing token embeddings.

\noindent\textbf{Interpolation Initialization} initializes the embeddings for the lowest and highest pitch tokens ($\langle 48 \rangle$ and $\langle 83 \rangle$) using Gaussian initialization. 
The embeddings for the pitches in between are linearly interpolated between these two.

\noindent\textbf{Scalar Initialization} begins by initializing a central pitch token $\langle 66 \rangle$ using Gaussian initialization. The embeddings for the remaining pitches are then set as multiples of this central pitch embedding, where multipliers range from $[-\ln(e+17), \cdots, -\ln(e), \ln(e), \cdots, \ln(e+17)]$. Compared to interpolation initialization, scalar initialization is more like a special form of extrapolation.


Empirically, we find that the scalar initialization works best for pitch modeling. For more details, please refer to the ablation study as shown in Table~\ref{tab:pitch}. Therefore, we use scalar initialization on pitch tokens for SongComposer.

\subsection{Progressive Structure-aware Training}
\label{sec:train}
Structure is crucial in song composition, with a typical song comprising multiple levels of structure~\citep{dai2022missing}. Therefore, we meticulously devise three stages of training for SongComposer to emphasize structural information at varying levels of time granularity.

\noindent\textbf{Motif-Level Melody Training.} 
Motif, in song composition, denotes a recurring musical idea that is key to enhancing the structure and coherence of the piece. Typically, a motif comprises a sequence of notes that repetitively appear throughout the song. Motivated by this concept, we intentionally select highly repetitive short 
note sequences to construct motif-level melody data. Subsequently, we kick off the training process of SongComposer by introducing this finely repetitive structure. In this way, the model is introduced to learn the motif-level composition.

\noindent\textbf{Independent Whole-Song Lyric and Melody Training.}
After gaining insight into the basic units of composition through motif-level melody training, we extend the training of SongComposer to the whole-song level. 
However, directly training the model to establish alignments between melody and lyrics may expose challenges to SongComposer.
Therefore, we continue to train the model using pure lyric and pure melody datasets to establish a foundation for basic whole-song understanding. 

\noindent\textbf{Paired Lyric and Melody with Phrase-level Token Training.}
Having a broader temporal dimension than a motif, the concept of phrases is also pivotal in structuring a song. A phrase is a sentence-level pattern that expresses a complete musical thought. To incorporate this understanding into composition, SongComposer trains on paired lyric and melody data, integrating the concept of phrases into the paired data. In our paper, we focus on five commonly used phrases such as `intro', `verse', `chorus', `bridge', and `outro', while unifying less common phrases as `other'.
Each phrase in a song would be outlined by two special tokens to indicate its beginning and end, resulting in a total of $6 \times 2$ special tokens. To maintain the model's ability to process melodies and lyrics separately, we train an equal amount of pure melody, pure lyric, and paired data. In contrast to the previous stage, both the pure melody and pure lyric data are now decorated with phrase-level special tokens.

\begin{table*}[h]
    \centering
    \small
    \begin{tabular}{l|lll| ccc}
    \bottomrule
    \multirow{2}{*}{Method} & \multicolumn{3}{c|}{Lyric-to-Melody} & \multicolumn{3}{c}{Melody-to-Lyric} \\ \cline{2-7}
        & PD(\%) $\uparrow$ & DD(\%) $\uparrow$ & MD $\downarrow$ & Cosine Dist. $\uparrow$ & ROUGE-2 $\uparrow$ & BS $\uparrow$ \\\hline
        SongMass~\cite{sheng2021songmass} & 30.34 & 48.98 & 2.95 & 0.568 & 0.204 & 0.532 \\ 
        TeleMelody~\cite{ju2022telemelody} & 46.81 & 51.77 & 2.60 & - & - & -\\\hdashline
        LLaMA 2~\cite{touvron2023llama} & 12.10 & 32.56 & 9.21 & 0.625 & 0.153  & 0.560\\
        InternLM 2~\cite{2023internlm} & 16.32 & 34.25 & 5.77 & 0.636 & 0.124 & 0.505\\
        Qwen 1.5~\cite{bai2023qwen} & 20.69 & 39.37 & 4.11 & 0.592 & 0.136 & 0.589 \\
        GPT-3.5~\cite{openai2022chatgpt} & 31.24 & 38.52 &  3.01 & 0.641 &  0.142 & 0.603 \\
        GPT-4~\cite{openai2023gpt4} & 36.43 & 42.94 & 2.87 &  0.654 &  0.158 & 0.610  \\\hdashline
        InternLM 2 + FT & 24.78 & 38.96 & 4.03 & 0.621 & 0.144 &  0.575\\
        SongComposer & \textbf{50.75} &\textbf{57.71} & \textbf{2.20} & \textbf{0.697} & \textbf{0.234} & \textbf{0.657} \\
    \bottomrule
    \end{tabular}
    \caption{Objective evaluation of Lyrics-to-Melody and Melody-to-Lyrics tasks. For open-source LLMs, we select models with a size of 7B parameters. As for GPT models, we utilize the most recent versions, namely gpt-4-turbo and gpt-3.5-turbo. InternLM 2 + FT stands for fine-tuning the InternLM 2 without incorporating any proposed techniques in this paper.}
    \label{tab:l2m}
\end{table*}

\section{Experiments}
\label{sec:exps}
\subsection{SongCompose Dataset}
To train the SongComposer, we curate a large-scale song pretraining and supervised fine-tuning dataset SongCompose. For more details, please refer to our Appendix~\ref{app:dataset}.

\noindent\textbf{Pure-lyric Dataset.}
We collect 283K song lyrics from two online sources, including 150K English lyrics and 133K Chinese lyrics. After a series of lyric-cleaning processes, we gather high-quality lyrics from various genres and styles.

\noindent\textbf{Pure-melody Dataset.}
We collect 20K MIDI files and extract melody attributes, including note pitch, note duration, and rest duration. We employ the \textit{pretty\_midi} Python module~\cite{raffel2014intuitive} to parse MIDI files and extract the "melody" or "vocal" tracks as the pure melody.

\noindent\textbf{Paired Lyric-melody Dataset.}
We create from scratch a dataset of 8K pairs of lyrics and melodies from the Internet, with roughly half being in Chinese and the other half in English. Melodies and lyrics are matched at the word level. 

\noindent\textbf{Supervised Finetuning Dataset.} We curate instruction-following data for song-generation tasks including creating melodies for given lyrics, writing lyrics for melodies, extending song segments, and generating songs from text descriptions. Specifically, we manually prepare 3K QA pairs for each of the first three tasks. Additionally, for the final task, we use GPT-4 to produce 1K song descriptions, which forms a text-to-song dataset that guides the song creation process.

\begin{table*}[]
    \centering
    \small
    \begin{tabular}{l|cc | cc | cc | cc}
    \bottomrule
        \multirow{2}{*}{Method} & \multicolumn{2}{c|}{Lyric-to-Melody} & \multicolumn{2}{c|}{Melody-to-Lyric} & \multicolumn{2}{c|}{Song Continuation} & \multicolumn{2}{c}{Text-to-Song} \\ \cline{2-9}
        &HMY.$\uparrow$ & MLC.$\uparrow$ & FLN.$\uparrow$ & MLC.$\uparrow$ & OVL.$\uparrow$ & COH.$\uparrow$ & OVL.$\uparrow$ & REL.$\uparrow$ \\\hline
        GPT-3.5~\cite{openai2022chatgpt} & 1.68 & 1.88 & 2.90 & 2.99 & 2.67 & 2.84 & 2.53 & 2.95\\
        GPT-4~\cite{openai2023gpt4} &2.82 &  2.79 & 2.84 & 3.20 & 2.86&  3.10& 2.43 & 3.27\\
        SongComposer &\textbf{3.82} &\textbf{3.76} &\textbf{3.63} &\textbf{3.69} &\textbf{3.61} &\textbf{3.58} &\textbf{3.41}  &\textbf{3.88}\\
    \bottomrule
    \end{tabular}
    \caption{Subjective evaluation of four tasks. Harmony (HMY.), Melody-Lyric Compatibility (MLC.), Fluency (FLN.), Overall Quality (OVL.), Coherence to Song Prompt (COH.), and Relevance to Text Input (REL.) depict the quality of each method in generating musically harmonious, lyrically coherent, and contextually relevant songs.}
    \label{tab:sub_eval}
\end{table*}

\begin{table*}[htbp]
\begin{minipage}[b]{0.5\linewidth}
    \centering
    \begin{tabular}{ll ll}
    \toprule
   Method & MD $\downarrow$ & BS $\uparrow$&  RR $\uparrow$ \\\midrule
    GPT-3.5~\citep{openai2022chatgpt}  &  2.88 & 0.601 & 1.13 \\
    GPT-4~\citep{openai2023gpt4}  & 2.73 & 0.613 & 1.25\\ 
    SongComposer$\dagger$  & 2.58 & 0.612 & 1.35\\
    SongComposer  & \textbf{2.12} & \textbf{0.662} & \textbf{1.64} \\
    \bottomrule
    \end{tabular}
    \caption{Objective evaluation of song continuation. $\dagger$ means the exclusion of phrase-level tokens.}
    \label{tab:song_continuation}
\end{minipage}
\hspace{0.1\linewidth}
\begin{minipage}[b]{0.4\linewidth}
    \centering
    \begin{tabular}{lcc}
    \toprule
        Init Method  & MD $\downarrow$ & RR $\uparrow$ \\\midrule
        Average & 3.07 & 1.44 \\
        Gaussian & 3.41 & 1.83 \\
        Interpolation & 2.90 & 1.77 \\
        Scalar & 2.33 & 2.03 \\\bottomrule 
    \end{tabular}
    \caption{Ablation study on pitch initialization methods.}
    \label{tab:pitch}
\end{minipage}
\end{table*}

\begin{table*}[htbp]
\small
\begin{minipage}[t]{0.55\linewidth}
    \centering
    \begin{tabular}{lllllll}
    \toprule
    Repeat threshold & 5 & 10 & 15 & 20  & 25 & $\infty$ \\
    Quantity & 12.3$\times$ & 4.6$\times$ & 2.8$\times$ & 1.5$\times$ & 1$\times$ & 0$\times$ \\ \midrule
    RR  $\uparrow$ &	2.19&	2.03&	1.61&1.58&	1.51&	1.45\\
    MD $\downarrow$ & 2.62 & 2.33	&3.25	&3.05	&4.48 & 3.07\\
    \bottomrule
    \end{tabular}
    \caption{Ablation on the repetition of motif-level melody data.}
    \label{tab:motif}
\end{minipage}
\hspace{0.05\linewidth}
\begin{minipage}[t]{0.4\linewidth}
    \centering
    \begin{tabular}{lll}
    \toprule
    Alignment & MD $\downarrow$ &  BS $\uparrow$ \\
    \midrule
    Song-Level& 3.71& 0.572 \\
    Line-Level& 2.42& 0.623 \\
    Word-Level& 2.12& 0.662 \\
    \bottomrule
    \end{tabular}
    \caption{Ablation on alignment granularity.}
    \label{tab:alignment}
\end{minipage}
\end{table*}

\subsection{Training Details}
\label{sec: training}
We adopt InternLM2-7B~\cite{cai2024internlm2} as our base model and set the maximum token length as 5120. Except for the pitch tokens using Scalar initialization, all the other newly added special tokens adopt Gaussian initialization.
We train the whole model to predict the next token based on prior text, maximizing the log-likelihood of tokens in the given examples. 
For optimization, we use AdamW optimizer~\cite{loshchilov2018decoupled} with a learning rate of $10^{-5}$, $\beta_1 = 0.9$, $\beta_2 = 0.95$, and a weight decay of 0.1. The entire dataset is iterated through once, with a batch size of 1. Additionally, a linear warm-up of the learning rate is applied during the initial $1\%$ of training steps, increasing from $10^{-6}$ to $10^{-5}$. Afterwards, a cosine schedule is applied, reducing the learning rate to a minimum of 0. This setting is consistent across both the pretraining and supervised fine-tuning stages. The whole training is conducted on 16 Nvidia A100 (80G) GPUs for approximately 2 days.

\subsection{Evaluation Setup}
We construct a validation set of 100 songs, evenly split between Chinese and English, none of which were seen by our model during training.

For objective metrics, we measure melody and lyric generation in three folds respectively.

\noindent\textbf{Melody Generation.}
For assessing the similarity between the generated melodies and the ground-truth, we adopt the metrics proposed by SongMASS~\cite{sheng2021songmass}: Pitch Distribution Similarity (PD), Duration Distribution Similarity (DD), and Melody Distance (MD). 
Besides, we propose a recall rate to assess the repetition capability and partly indicate the structure within the song. This rate is calculated by dividing the total number of melodic lines by the number of unique melodic lines, with a minimum recall rate of 1 indicating no repetition.


\noindent\textbf{Lyric Generation.}
We evaluate the similarity between generated and original lyrics using three metrics from different perspectives. We use a CoSENT (Cosine Sentence) model~\cite{text2vec}, specifically the base-multilingual version, to compute sentence-level cosine similarity. Additionally, we apply the ROUGE-2 score~\cite{lin2004rouge} to measure bigram overlap and the BERT score (BS)~\cite{zhang2019bertscore} to assess similarity based on the contextual embeddings from the BERT-base model.


For the subjective evaluation, we conduct a user study in which 30 participants assessed 10 instances for each task. We develop two metrics for each task and ask the participants to rate them. The rating scale is $1$ to $5$, where higher scores denote superior quality. More details can be found in Appendix~\ref{app:human}. In this way, we collect feedback on the quality of the generated content from a human perspective.

The evaluation criteria for different tasks are as follows:
For Lyric-to-Melody Generation, we assess Harmony and Melody-Lyric Compatibility.
For Melody-to-Lyric Generation, we evaluate Fluency and Melody-Lyric Compatibility.
Song Continuation quality is assessed based on Overall Quality and Coherence to the Song Prompt.
Text-to-Song generation is evaluated in terms of Overall Quality and Relevance to the Text Input.

In summary, each task is evaluated using two metrics: one that assesses the overall musical quality of the samples produced, and another that specifically addresses the challenges of each task. More detailed descriptions of the tasks and metrics can be found in Appendix~\ref{app:sub}.

\subsection{Experimental Results}

We test and compare our method majorly with existing LLMs.
For the alternative LLM baselines, we employ a few-shot prompt approach, feeding sample examples to prompt the LLM and produce the desired output following the given instructions. Details are provided in the Appendix~\ref{baseline_appendix}.
We gather outputs from both GPT-4 and GPT-3.5 via their APIs. Additionally, we also assess other typical LLMs whose weights have been obtained from the Hugging Face community. 

\noindent\textbf{Objective Evaluation.}
Table~\ref{tab:l2m} presents a comparison of methods for converting lyrics to melody and vice versa. Compared to traditional methods, which include special designs for specific tasks, SongComposer still shows significant improvement.
SongComposer significantly outperforms advanced large language models such as GPT-4 in both the lyric-to-melody task and the melody-to-lyrics task.
Moreover, simple fine-tuning on InternLM 2 does not produce rational melodies and lyrics, showing the effectiveness of our systematic design. As shown in Table~\ref{tab:song_continuation}, SongComposer excels not only in generating high-quality lyrics and melodies individually but also in jointly producing both by continuing given lines. Since there is no objective evaluation for text-to-song generation, we provide a formatted musical score in Appendix~\ref{sec:case_study}. The song generated by SongComposer is well-structured and coherent to the prompt.


\noindent\textbf{Subjective Evaluation.}
The subjective evaluation in Table~\ref{tab:sub_eval} highlights that SongComposer significantly surpasses GPT-3.5 and GPT-4 in overall quality, coherence to the prompt, and melody-lyric compatibility. This underscores SongComposer's advanced capability to capture the song's structure and generate a harmonized melody and lyrics that seamlessly fit together.

\subsection{Ablation Study}
In the ablation study, we probe the SongComposer on song continuation task and report the Melody Distance (\textbf{MD}), Recall Rate (\textbf{RR}), and BERT Score (\textbf{BS}) to depict the quality of melody and lyric respectively. All studies are conducted on the validation set, except for the memorization analysis, where we use the training data to test whether the model memorizes the training set.

\noindent\textbf{Phrase-level Special Tokens.}
To study the importance of phrase-level indication in song composition, we train the SongComposer model without phrase-level special tokens. The results, presented in Table~\ref{tab:song_continuation}, show a significant decline in generation quality when these tokens are omitted. Moreover, the use of phrase-level special tokens improves the model's ability to capture recurring musical ideas, as evidenced by an increase in the recall rate. Both observations suggest that phrase-level indications are essential for producing coherent and fluid song compositions.

\begin{figure*}[t]
    \centering
    \begin{minipage}{0.45\textwidth}
        \centering    
        \includegraphics[height=0.2\textheight, keepaspectratio]{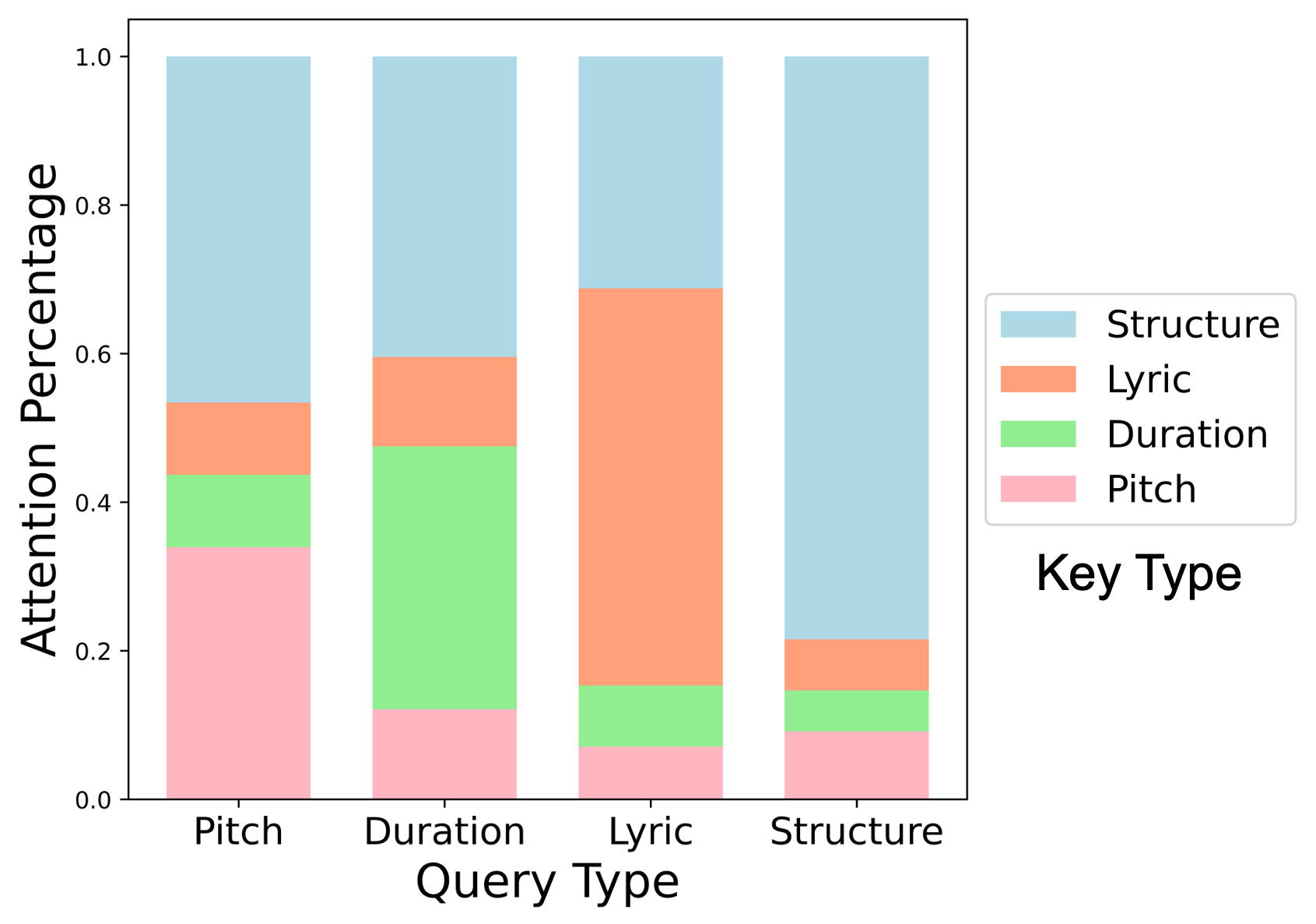}
        \caption{Visualization of attention distribution for different key/query types.}
        \label{fig:attn_dist}
    \end{minipage}%
    \hspace{0.08\textwidth}
    \begin{minipage}{0.45\textwidth}
        \centering
        \includegraphics[height=0.2\textheight, keepaspectratio]{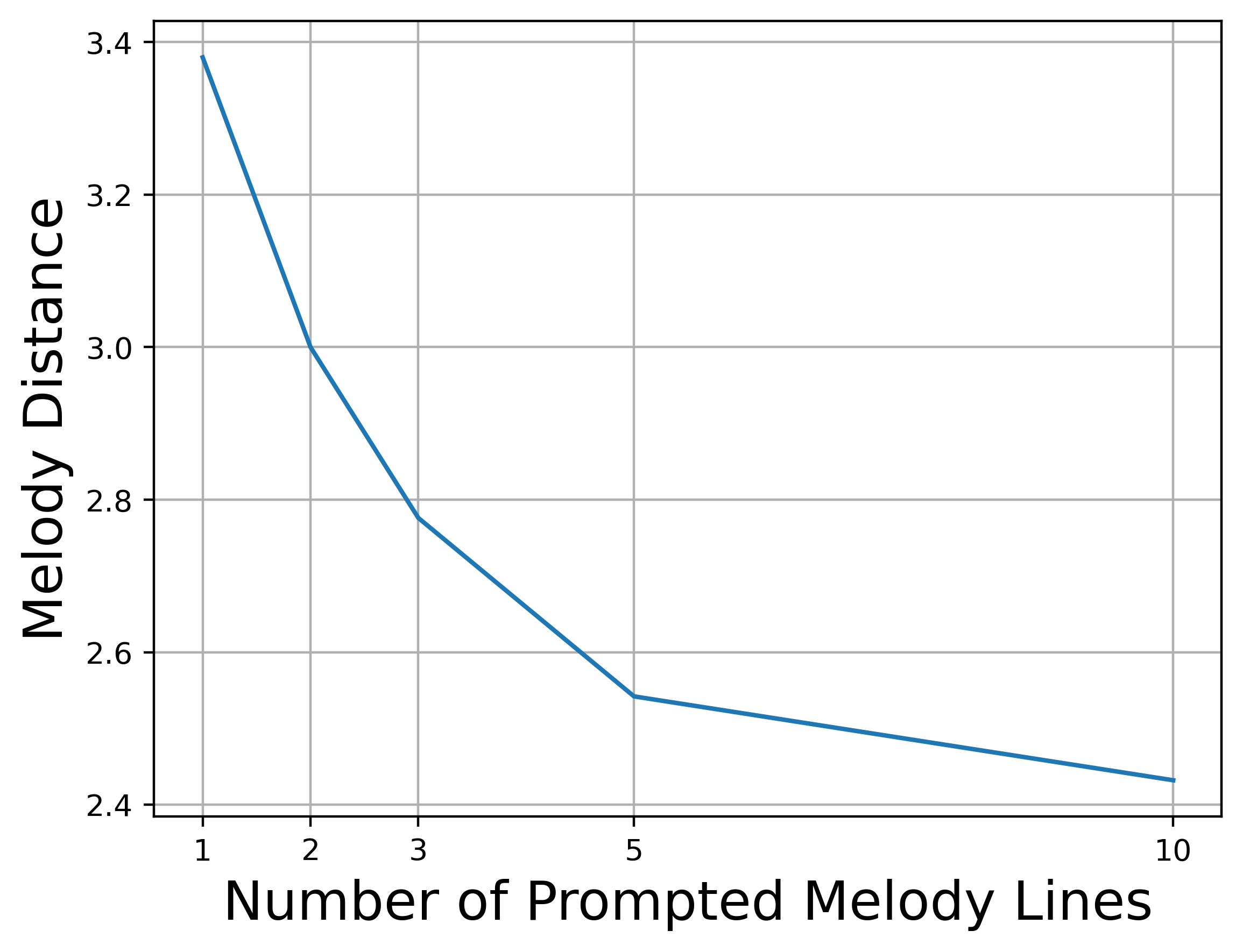}
        \caption{Memorization analysis of SongComposer.}
        \label{fig:mem}
    \end{minipage}
\end{figure*}

To delve deeper into the influence of phrase tokens and the interplay between musical elements during generation, we categorize the input into four primary types: lyric, duration, pitch, and structure. The structure type here refers to phrase-level tokens. We then analyze the attention maps from all layers of SongComposer. The attention distribution shown in Figure~\ref{fig:attn_dist} reveals that structural phrase-level tokens have a profound impact across all query types, showing the crucial role of structure in song generation. Furthermore, the model tends to prioritize musical elements that are consistent with the input query's type. For instance, when processing lyric queries, the model allocates nearly half of its attention to keys related to lyrics.

\noindent\textbf{Pitch Initialization.}
We evaluate these four methods on the pure melody continuation task. As shown in Table~\ref{tab:pitch}, the scalar initialization presents a considerable advantage over other methods. We conjecture that the scalar method provides a strong prior on both the magnitude and direction of the newly initialized embeddings, which induces the model to learn the pitch patterns comprehensively.

\noindent\textbf{Motif-level Melody Data.}
To determine whether motif-level patterns enhance melody generation, we conduct baseline experiments where we train SongComposer exclusively on a pure melody dataset. We then test melody continuation, reporting melody distance (MD) and recall rate. We adjust the repetition threshold to control the level of motif repetition in the data. The repeat threshold refers to the frequency with which a motif appears within the melody dataset. A higher repeat threshold indicates a more common motif. For example, if the repetition threshold is set to 10, only motifs that appear more than 10 times in the dataset are included. The results are presented in Table~\ref{tab:motif}, with the baseline result in the right-most column where no motif-level data is inputted.

Firstly, all results with motif-level data boost the baseline in terms of recall rate, aligning with our intuition that injecting motif-level data improves the structure awareness of melody composition. Secondly, we find that a small amount of highly repetitive motif-level data can hurt the melody generation. We conjecture this is because highly repetitive motifs lack diversity and trap the model in a constrained generation space. Then, the melody distance reaches an optimal point at a threshold of 10, suggesting that a moderate degree of repetition achieves the best balance between motif variety and overall repetitiveness. Therefore, we extract motif-level melodies by a repeat threshold of 10.

\noindent\textbf{Pair Alignment at Different Granularity.}
We explore three methods for integrating lyrics and melody into a cohesive format. First, the \textbf{song-level} approach concatenates the entire set of lyrics for a song and the complete melody for that song.
Second, the \textbf{line-level} method connects each line of lyrics with the corresponding line of melody.
Finally, the \textbf{word-level} method merges each individual word of the lyrics with a single note of the melody. We provide an example of the word-level pairing format in Section \ref{sec:format} and illustrate the other two alignment methods in Appendix~\ref{app:align_example}.

Table~\ref{tab:alignment} shows that finer alignment improves generation quality. Word-level alignment has the lowest melody distance and highest BERT score, indicating the best performance. Furthermore, we observe that the song-level and line-level pairing formats often fail to accurately produce the corresponding melody and lyrics in terms of quantity, thereby diminishing the overall generation quality.

\noindent\textbf{Memorization analysis on SongComposer.}
To investigate the extent to which SongComposer memorizes the training data, we conduct a memorization analysis inspired by MusicLM \cite{agostinelli2023musiclm}. Specifically, we prompt training melody data samples with varying numbers of lines and compare the generated melodies to their original target counterparts. We quantify the similarity between the two melodies using melody distance, which would approach 0 if the prediction exactly matches the target.
As shown in Figure~\ref{fig:mem}, we find that the melody distance remains relatively high even when prompted with 10 lines of melody, indicating that our strategy is not trivially memorizing the training data and the generated results differ from the corresponding sequences in the train set.

\section{Conclusion}
This paper introduces SongComposer, a novel large language model for generating music scores that synchronize lyrics and melodies. The model uses a tuple format to align lyrics and notes at the word level and employs scalar initialization for note pitch, enhancing pitch modeling efficiency. A multi-stage pipeline is implemented during training, capturing structure from motif-level melody data to phrase-level indicators for better coherence and repetition. Our experiments demonstrate that SongComposer outperforms traditional methods and models like GPT-4, showcasing its potential for music creation.

\section*{Limitations}
\label{sec:limit}
SongComposer primarily focuses on generating symbolic music that synchronizes lyrics and melodies. However, producing corresponding audio currently requires supplementary singing voice synthesis tools. While the musical quality of the audio is partly dependent on the generated score (SongComposer's output), it also significantly relies on the singer's performance, including timbre and vocal techniques—areas outside the scope of symbolic music generation. This distinction is important for evaluating the performance of our model, as the perceived audio quality is heavily influenced by the synthesis tool used, not solely by our work. Additionally, integrating multi-track accompaniment generation and expressive performance modeling could further enhance the system's capability.

Symbolic music generation offers fine control and superior editability, whereas acoustic methods provide impressive musical expressiveness and listenability. In future work, we aim to integrate symbolic and acoustic approaches to create full-track songs. This integration will enable the generation of precise scores alongside their corresponding high-quality audio, achieving a balance between control and auditory appeal.

\section*{Ethics Statements}
\label{sec:impact}
The proposed work, SongComposer, a large language model designed for generating songs, has the potential impact on various aspects of society. On the positive side, SongComposer effortlessly creates high-quality songs with melodies and lyrics which can optimize the music creation process, allowing individuals with limited musical training to express their creativity and contribute to the music landscape.

However, as SongComposer generates songs autonomously, there is a risk of potential copyright infringement or misuse of intellectual property. We have conducted a preliminary memorization analysis shown in Figure~\ref{fig:mem}. However, proper measures still need to be in place to ensure that the generated songs adhere to copyright laws and protect the rights of original composers and authors.

In conclusion, while SongComposer presents exciting possibilities for the music industry and creative expression, its development should be accompanied by careful consideration of ethical and societal implications.

\section*{Acknowledgement}
This work was supported by National Key R\&D Program of China 2022ZD0161600, Shanghai Artificial Intelligence Laboratory, the Centre for Perceptual and Interactive Intelligence (CPII) Ltd under the Innovation and Technology Commission (ITC)’s InnoHK. Dahua Lin is a PI of CPII under the InnoHK.




\bibliography{custom}

\newpage
\appendix

\section{SongCompose Dataset}
\label{app:dataset}
This section introduces the compilation, creation, and statistical breakdown of our SongCompose dataset, which includes separate collections of lyrics, melodies, and lyric-melody pairs that synchronize lyrics with melodies at the word level. We aim to publicly release this three-fold dataset and following supervised fine-tuning dataset, providing a foundational resource for future research. 
\subsection{Pure-lyric Dataset}
We collect pure lyrics datasets from two online sources:
(1) The Kaggle dataset\footnote{\url{https://www.kaggle.com/datasets/edenbd/150k-lyrics-labeled-with-spotify-valence}}, comprising the lyrics of 150K songs labeled with Spotify Valence, a measure of the positiveness of the song.
(2) The Music Lyric Chatbot dataset\footnote{\url{https://github.com/liuhuanyong/MusicLyricChatbot}}, containing the lyrics of 140K Mandarin-language songs.
After a series of lyric-cleaning processes, we gather high-quality lyrics from 283K songs, including 150K in English and 133K in Chinese. 

Table~\ref{tab:pure_lyrics} provides a detailed breakdown of the dataset, including language distribution, average lines per song, words per line, and the count of unique words. 

 \begin{table*}[h]
    \centering
    \begin{tabular}{ccccc}
    \toprule
    Language &\#Song  &\#Line/\#Song & \#Word/\#Line &\#Unique word  \\
    \midrule
    English & 150359 &34.4 &5.8 &168669\\
    Chinese  &132930 &28.0  &8.7 &7740 \\
    \midrule
    Total &283289 &31.4 &7.0 &176399 \\
    \bottomrule
    \end{tabular}
    \caption{Statistical details of the pure-lyric dataset.}
    \label{tab:pure_lyrics}
\end{table*}

\subsection{Pure-melody Dataset}
To organize the melody dataset into a text-based structure, we collect royalty-free MIDI files from various websites, e.g., \textit{midiworld} and \textit{freemidi}. Using MIDI files for our pure melody dataset offers inherent structural simplicity, enabling efficient extraction and manipulation of melodies without complex audio processing.   
Among our collection, 45K entries come from the LMD-matched MIDI dataset~\cite{Raffel2016LearningBasedMF}, while approximately 80K are acquired through web crawling.

For parsing MIDI files, we employ \texttt{pretty\_midi}~\cite{raffel2014intuitive}, a Python module designed for creating, manipulating, and analyzing MIDI files. We extract the "melody" or "vocal" tracks from these MIDI files. Since melody in MIDI is represented as a sequence of musical notes over time and each note has a specific pitch, start and end timestamp, we obtain a list of melody attribute triplets consisting of \{\textit{note pitch}, \textit{note duration}, \textit{rest duration}\}.

\begin{itemize}[itemsep=3pt,topsep=0pt,parsep=0pt]
    \item \textit{Note pitch:} The pitch of notes is represented by their corresponding MIDI note numbers, ranging from 0 to 127, with the number 60 predefined as Middle C.
    \item \textit{Note duration:} A note's duration is defined as the length of time in seconds that the note is played. This is computed from the start and end times of each note embedded within the MIDI files as follows: $\text{note-duration}_k = \text{note-end}_k - \text{note-start}_k$, where $k$ represents the note index number.
    \item \textit{Rest duration:} The rest duration represents the silent period that follows the playing of a note. It can be calculated by $\text{rest-duration}_k = \text{note-start}_{k+1} - \text{note-end}_k$.
\end{itemize}

We perform necessary data filtering to remove duplicate and poor-quality samples, leaving approximately 20K MIDI samples remaining.

\subsection{Paired Lyric-melody Dataset}

\begin{figure}[t!]
    \centering
    \includegraphics[width=\columnwidth]{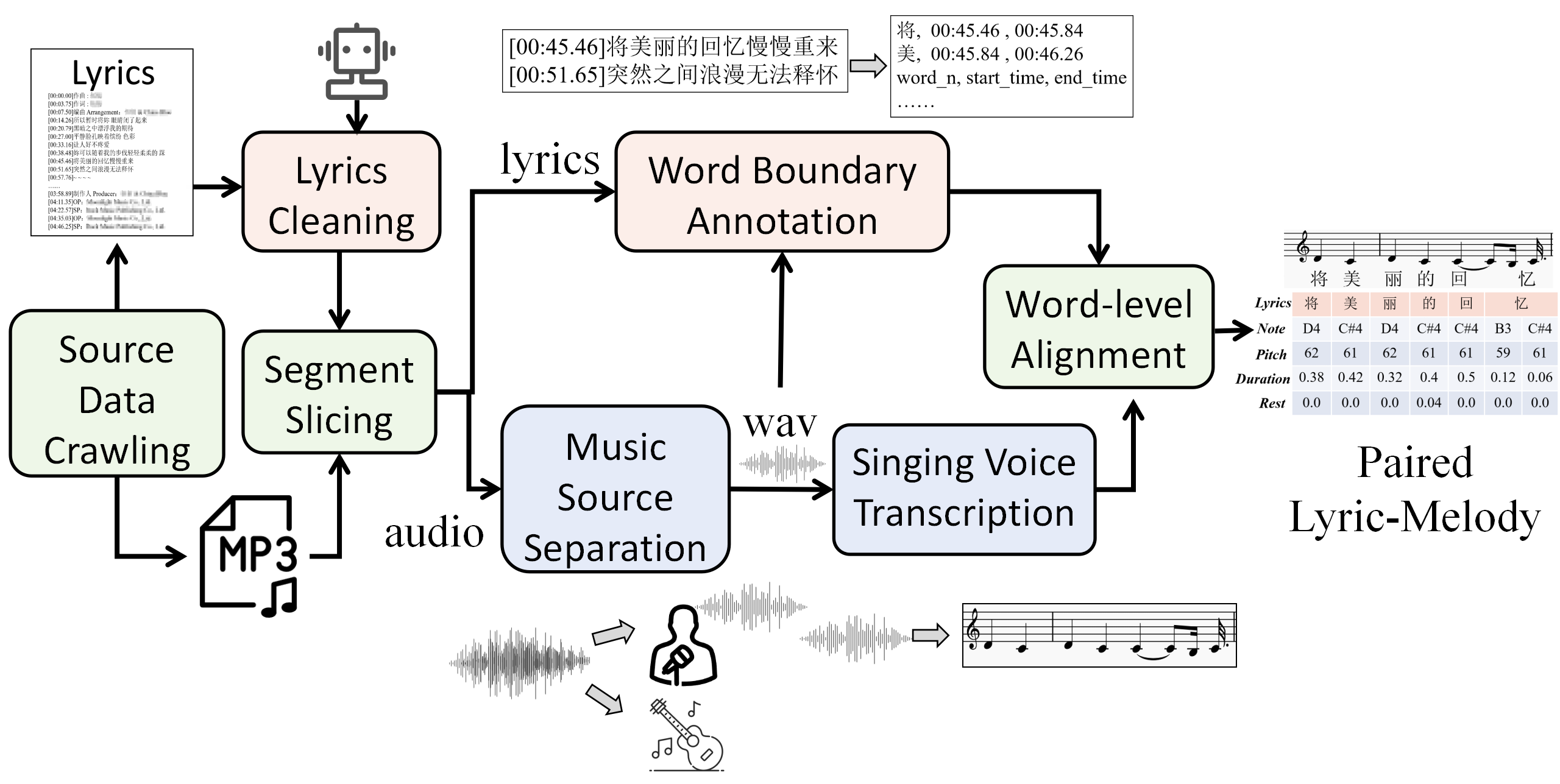}
    \caption{Pipeline of paired lyric-melody data collection.}
    \label{fig:pipeline}
\end{figure}

To build paired data with precise alignment, We efficiently process web-scraped information on a large scale to create a paired dataset of 4K classic Chinese songs and 4K English songs, gathering vocal data from copyright-free websites like the Free Music Archive and ccMixter. As illustrated in Figure~\ref{fig:pipeline}, the pipeline for collecting lyric-melody data is as follows:
\begin{enumerate}[itemsep=3pt,topsep=0pt,parsep=0pt]
    \item[(1)] Source Data Crawling: We crawl the web to gather a large dataset of mp3 files and their corresponding lyric files, encompassing sentence-level timestamps.
    \item[(2)] Lyrics Cleaning: We use GPT-4 to clean irrelevant details from lyric texts, such as song titles, artist names, and production information.
    \item[(3)] Segment Slicing: To mitigate the challenges and error accumulation for long-time alignments, we slice the audio and lyrics into paired segments of approximately 10 seconds (roughly three sentences each) based on timestamps provided in the lyric files. 
    \item[(4)] Music Source Separation: We utilize UVR\footnote{\url{https://github.com/Anjok07/ultimatevocalremovergui}}, a public music separation tool, to separate the vocal from the accompaniment part in the original audio.
    \item[(5)] Singing Voice Transcription: Using a singing voice wav input, we leverage an internal model to automatically generate the preliminary musical score, capturing note pitch and start-end times of each note. 
    \item[(6)] Word Boundary Annotation: We obtain the boundaries of each word in lyrics with an audio alignment tool, Montreal Forced Aligner\footnote{\url{https://github.com/MontrealCorpusTools/Montreal-Forced-Aligner}}.
    \item[(7)] Word-level Alignment: The dynamic time warping (DTW)~\cite{muller2007dtw} algorithm is utilized to align words and notes based on start-end times.
\end{enumerate}

For information at the phrase level, we use the All-In-One music structure analyzer~\cite{taejun2023allinone} to extract it.
Finally, we develop a dataset comprising 8K paired lyric-melody entries, with approximately 4K in Chinese and 4K in English.

\begin{figure*}[h]
    \centering
    \includegraphics[width=\textwidth]{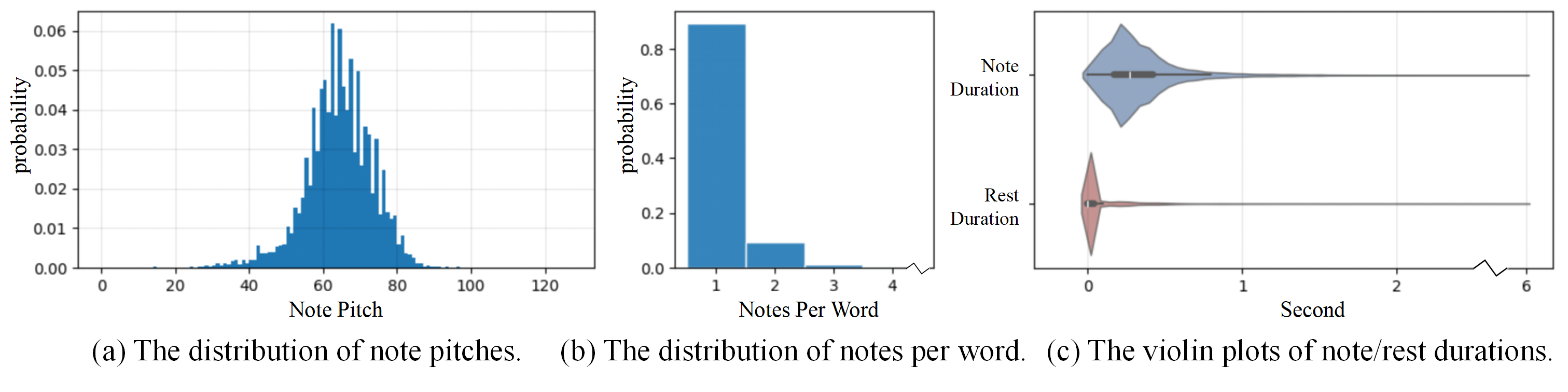}
    \caption{Distribution of music attributes in our paired lyric-melody dataset.}
    \label{fig:paired_distr}
\end{figure*}

We also conduct the statistical analysis of the paired lyric-melody dataset shown in Figure~\ref{fig:paired_distr}. We find that most pitch numbers fall within the range of 50 to 80 and the majority of words are paired with a single note, and around 10\% of words correspond to two or more notes. When examining note durations, we observe that they primarily vary between 0 to 1 second, and durations of rests are predominantly zero, reflecting a concise musical structure. 

\subsection{Supervised Finetuning Data}
\begin{figure*}[h]
	\centering
	\includegraphics[width=0.9\textwidth]{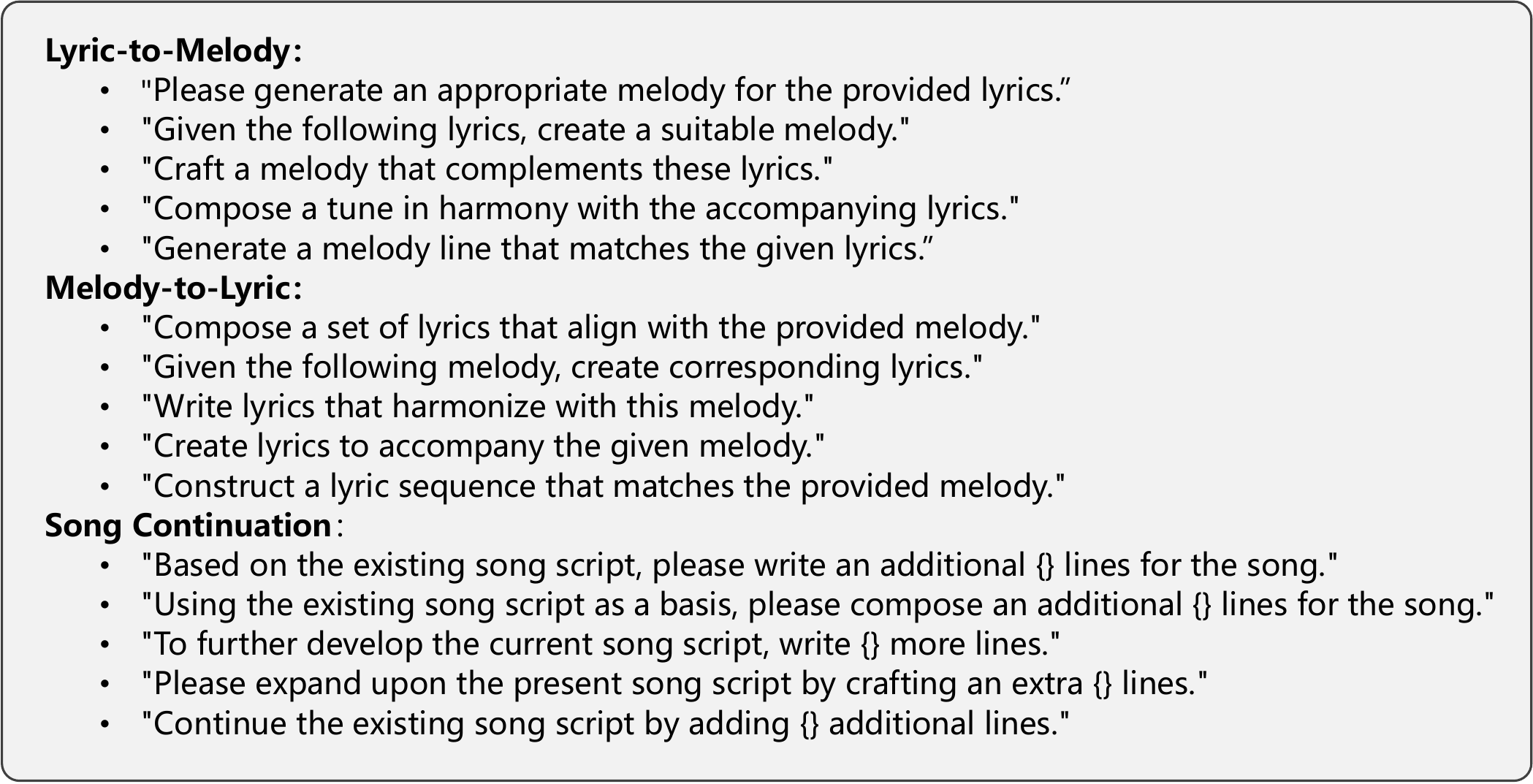}
	\caption{Instruction prompt templates for lyric-to-melody, melody-to-lyric, and song continuation tasks.}
	\label{fig:tasks_prompt}
\end{figure*}

\begin{figure*}[h]
	\centering
	\includegraphics[width=0.9\textwidth]{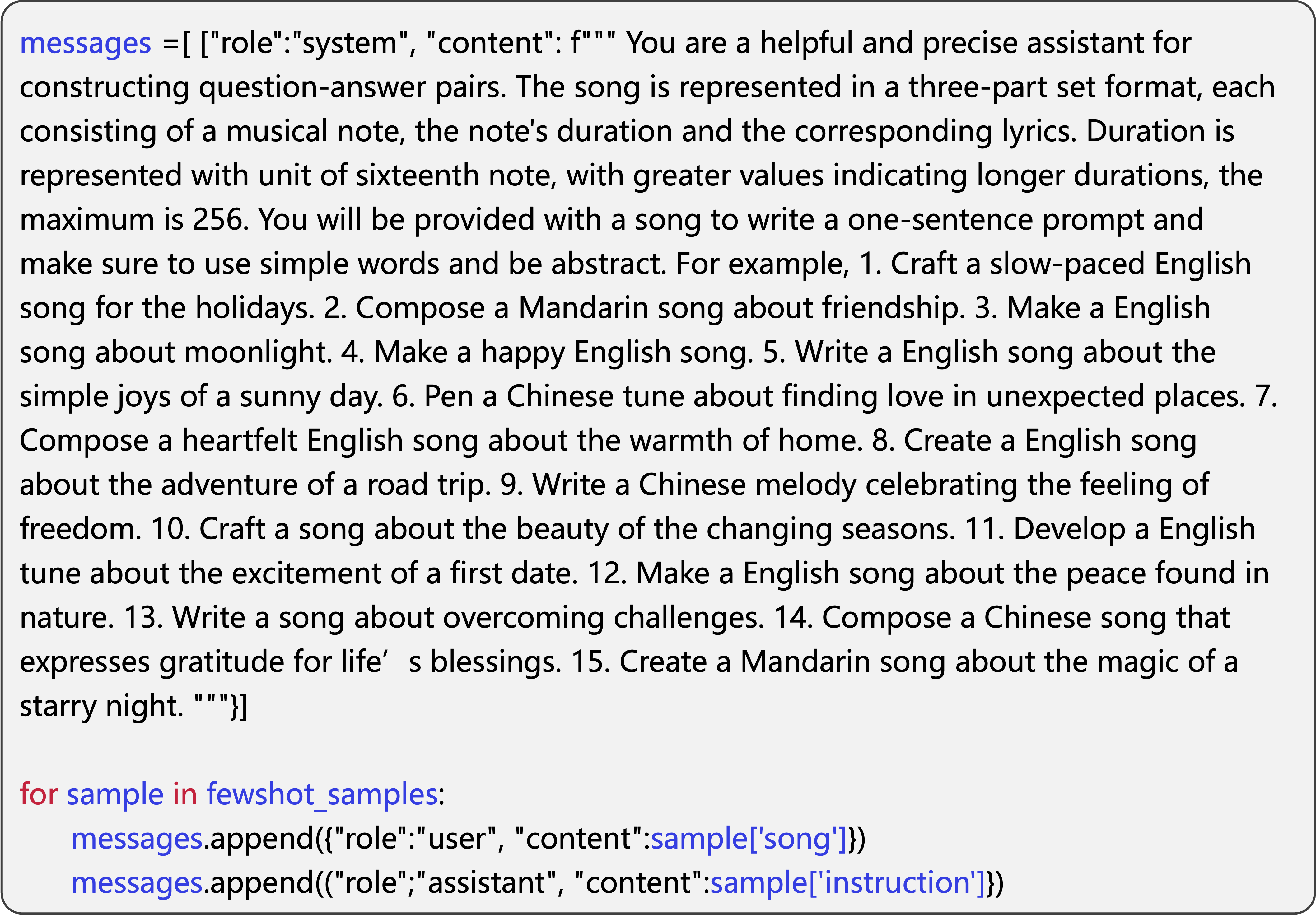}
	\caption{The prompt on text-to-song dataset construction process for GPT-4, using few-show in-context learning instructions.}
	\label{fig:text_song}
\end{figure*}

To achieve the instruction-following capability, we create supervised fine-tuning data for SongComposer. For lyric-to-melody, melody-to-lyric, and song continuation tasks, we manually design the prompt templates in Figure~\ref{fig:tasks_prompt}, which serve as the foundation for compiling our QA pairs. For example, in the lyric-to-melody task, we start with the instruction prompt, such as "Please generate an appropriate melody for the provided lyrics." Then the pure-lyric version of a song follows the prompt. The response then utilizes the lyric-melody paired version of the song. For the song continuation task, we will additionally specify the number of lines by which we want the model to extend the song.
To create the dataset for the final text-to-song task, we leverage the GPT-4 API. We feed the paired song data into the model and ask it to generate a prompt. We use a few-shot template to guide the output, as shown in Figure~\ref{fig:text_song}. Therefore, we can compile the text-to-song instruction set. 

\section{Details on Song-related Generation Tasks and Subjective Metrics}
\label{app:sub}
\noindent\textbf{Lyric-to-Melody Generation} asks to create a fitting melody based on the given lyrics. The melody is assessed on: 
(1) Harmony (HMY.): Evaluates the overall quality of the melody.
(2) Melody-Lyric Compatibility (MLC.): Examines how well the generated melody fits the given lyrics.

\noindent\textbf{Melody-to-Lyric Generation} aims to produce lyrics that match a provided melody. The lyrics are evaluated on:
(1) Fluency (FLN.): Considers the grammatical correctness and semantic coherence of the generated lyrics.
(2) Melody-Lyric Compatibility (MLC.): Examines how well the generated lyrics fit the given melody.

\noindent\textbf{Song Continuation} involves extending a given song segment both melodically and lyrically. We evaluate the continuation quality on:
(1) Overall Quality (OVL.): Measures the overall quality of the generated song in terms of its musical appeal.
(2) Coherence to the Song Prompt (COH.): Analyzes the natural integration of the continuation with the provided song prompt, assessing coherence in melody, lyrics, and other musical elements.

\noindent\textbf{Text-to-Song Generation} generates a complete song based on textual description, capturing its essence musically and lyrically. The evaluation focuses on:
(1) Overall Quality (OVL.): Measures the overall quality of the generated song in terms of its musical appeal.
(2) Relevance to the Text Input (REL.): Examines how well the generated song aligns with and derives relevance from the input text.

\section{Details on Human Evaluation}
\label{app:human}
\noindent\textbf{Participants' Musical Background}. We recruit 30 participants, all of whom have formal music education backgrounds and relevant musical training (e.g., independent music producers, music school teachers and students, band members), ensuring that the evaluations come from knowledgeable perspectives.

\noindent\textbf{Language Proficiency.} All participants are native Chinese speakers fluent in English, with strong abilities to understand and interpret lyrics in both languages.

\noindent\textbf{Demographic Distribution.} Participants ranged in age from 20 to 30 years old, evenly split by gender (50\% female, 50\% male). All participants hold at least a bachelor's degree or are currently undergraduate students.

\noindent\textbf{Recruitment Process.} All participants are volunteers without financial incentives.

\noindent\textbf{Task Description and Training.} Participants receive written instructions that include detailed examples and clearly defined evaluation criteria. They evaluate the generated samples based on two key dimensions: overall musical harmony and specific task-related challenges outlined in our study (see Appendix B for details). All evaluations are submitted through an online platform for efficient data collection.

\noindent\textbf{Evaluation Criteria and Scale}. Participants rate the results on a scale from 1 to 5, defined explicitly as follows: 1 - very poor, 2 - poor, 3 - average, 4 - good, and 5 - excellent.

\noindent\textbf{Sample Selection Strategy.} For each of the four tasks, we systematically select 10 test cases, ensuring coverage across various musical genres such as pop, rock, and ballad. Furthermore, half of the test cases are in English, while the other half are in Chinese.

\section{More Information on Ablation Study}
\subsection{Visualization on Pitch Initialization}
\label{app:vis_pitch}
\begin{figure*}[!h]
    \centering
    \includegraphics[width=0.99\linewidth]{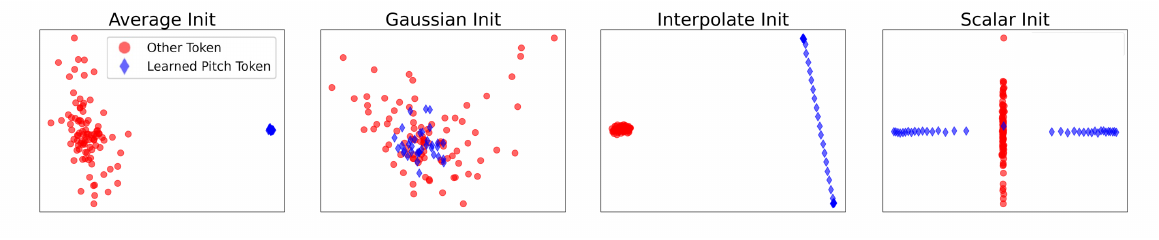}
    \caption{The visualization of learned pitch tokens and other tokens with different initialization methods. We use Principal Component Analysis (PCA) to reduce the dimensionality of the embeddings to 2 dimensions.}
    \label{fig:pitch}
\end{figure*}

To further interpret the learned pitch tokens under different initialization methods, we visualize the embeddings in Figure~\ref{fig:pitch}. 
The average initialization distinguishes between pitch and rest tokens but fails to capture the inherent pitch information, resulting in a collapsed cluster. 
The Gaussian method fails to differentiate between pitch tokens and other tokens effectively and does not learn a discernible pattern. 
For the remaining two methods, both initialization methods result in distinct patterns. The interpolation method positions pitch tokens far away from other tokens, while the scalar method results in a pattern where the mean cluster still lies among the existing tokens. Therefore, scalar initialization stays closer to the existing semantic spaces which may lead to a better generation than interpolation method.

\subsection{Example on Song-level and Line-level Pair Format}
\label{app:align_example}
In practice, the input of the song-level paired melody is formatted as follows:
\begin{align*}
``&\langle\text{bop}\rangle \langle\text{bol}\rangle \text{ Chinese/English song. Total } \{num\} \text{ lines.} \\
&\text{The 1-st line:} \, w_1 \, w_2 \, \cdots \\
&\text{The 2-nd line:} \, \cdots \, \langle\text{eol}\rangle \\
&\langle\text{bom}\rangle \text{ bpm is } \{bpm\}. \text{ Total } \{num\} \text{ lines.} \\
&\text{The 1-st line:} \, \langle p_1 \rangle, d_1 \,|\, \langle \text{rest} \rangle, r_1 \,|\, \langle p_2 \rangle, d_2 \,|\, \langle \text{rest} \rangle, r_2 \cdots\\
&\text{The 2-nd line:} \, \cdots \, \langle\text{eom}\rangle \langle\text{eop}\rangle"
\end{align*}

The input of the line-level paired melody is formatted as follows:
\begin{align*}
``&\langle\text{bop}\rangle \text{ Chinese/English song. } \text{ bpm is } \{bpm\}. \\ 
&\text{ Total } \{num\} \text{ lines.} \\
&\text{The 1-st line:} \, \langle p_1 \rangle, d_1 \,|\, \langle \text{rest} \rangle, r_1 \,|\, \\ 
&\langle p_2 \rangle, d_2 \,|\, \langle \text{rest} \rangle, r_2 \cdots || w_1 \, w_2 \, \cdots \\
&\text{The 2-nd line:} \, \cdots \, \langle\text{eop}\rangle"
\end{align*}

\begin{table}[]
    \centering
    \small
    \begin{tabular}{ccll}
    \toprule
    Pure-lyric & Pure-melody & MD $\downarrow$ & BS $\uparrow$ \\\hline
     \xmark  & \xmark   &  2.59 &  0.603 \\
    \xmark  & \cmark  & 2.29 & 0.618  \\ 
    \cmark & \xmark   & 2.44 & 0.645   \\
    \cmark & \cmark & 2.12 &  0.662 \\
    \bottomrule
    \end{tabular} 
    \vspace{5pt}
    \caption{Ablation study on pretraining datasets. \xmark\ denotes the exclusion of a specific dataset, while \cmark\ indicates its inclusion in the training. Paired data are used in all settings.}
    \label{tab:ablation_dataset}
\end{table}

\subsection{Ablation on Independent Lyric and Melody Training}
To explore the impact of specialized datasets on our model's learning, we conduct training experiments using paired data combined with different pure-lyric and pure-melody datasets. Table~\ref{tab:ablation_dataset} demonstrates that omitting both the pure-lyric and pure-melody datasets significantly reduces performance, highlighting the critical importance of foundational melodic and lyrical knowledge in the training stages.

Integrating either dataset individually results in notable improvements across tasks. Specifically, the pure-lyric dataset mainly enhances performance in the lyric-related generation, while the pure-melody dataset significantly boosts melody generation. This finding aligns with the intuitive understanding that each dataset enhances the model's comprehension of its respective modality.
Moreover, using both types of datasets together yields the best results, demonstrating a synergistic effect. 

\section{Tuple Format Examples}
\label{app:tuple}
\begin{figure*}[t!]
	\centering
	\includegraphics[width=0.9\textwidth]{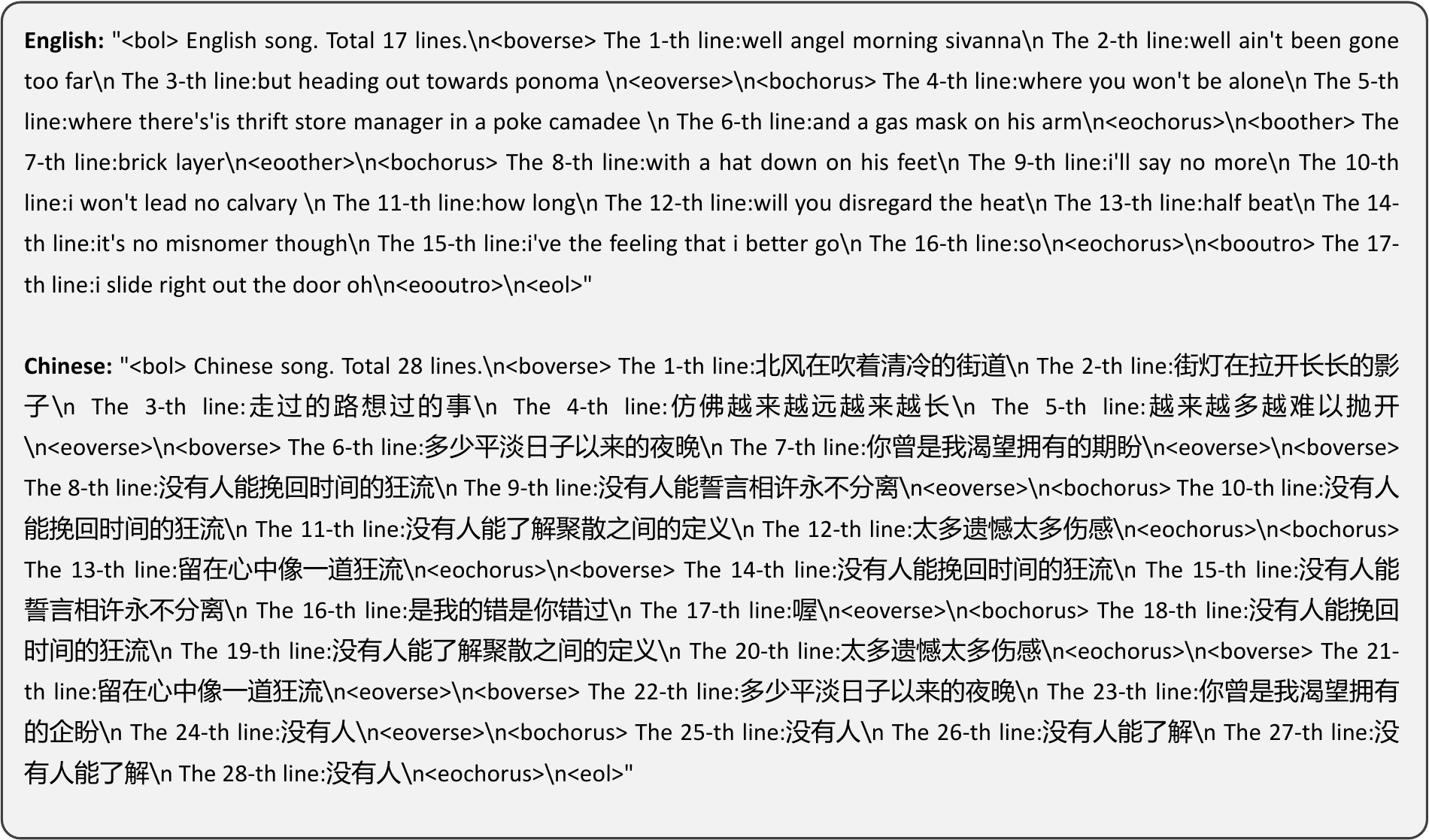}
	\caption{Two examples of lyric data in English and Chinese with phrase-level tokens.}
	\label{fig:lyric_case}
\end{figure*}

\begin{figure*}[t!]
	\centering
	\includegraphics[width=0.9\textwidth]{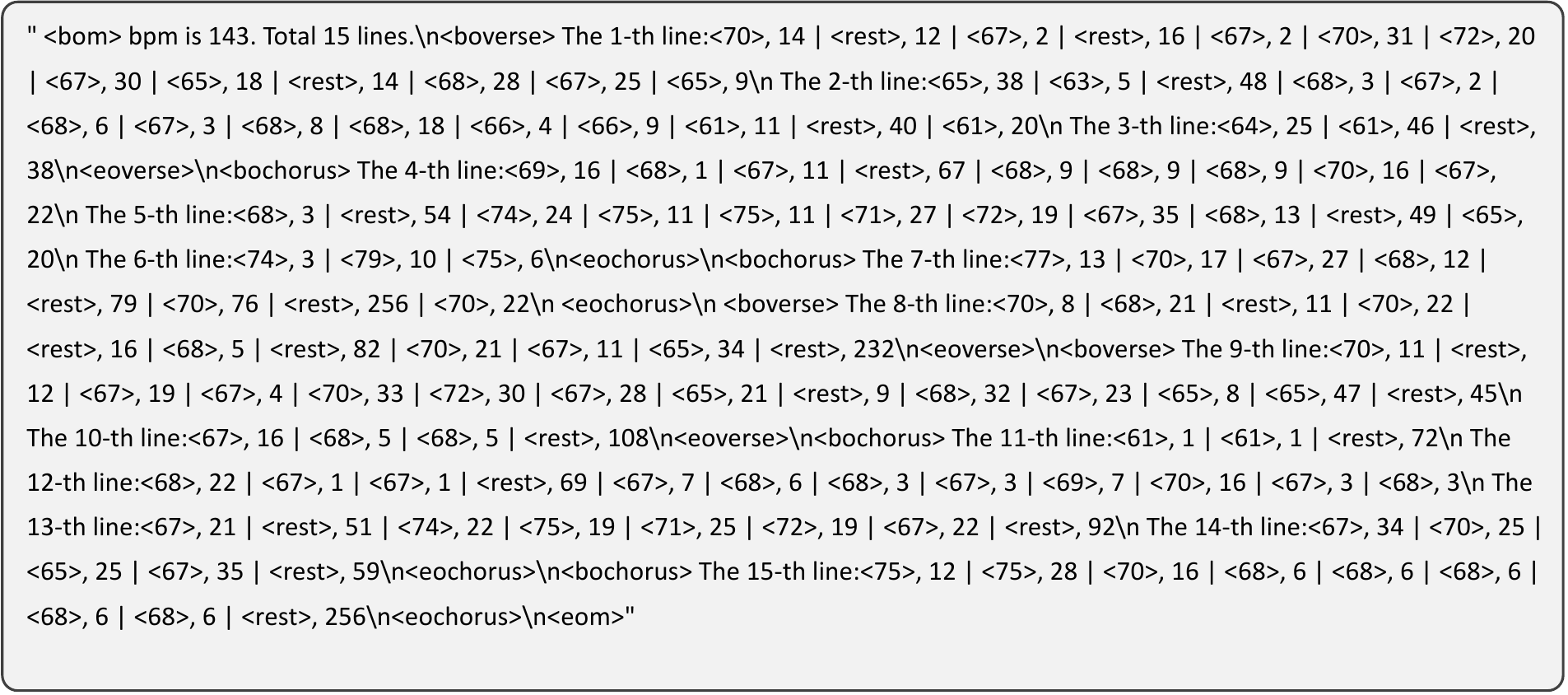}
	\caption{An example of pure melody data with phrase-level tokens.}
	\label{fig:melody_case}
\end{figure*}

\begin{figure*}[t!]
	\centering
	\includegraphics[width=0.9\textwidth]{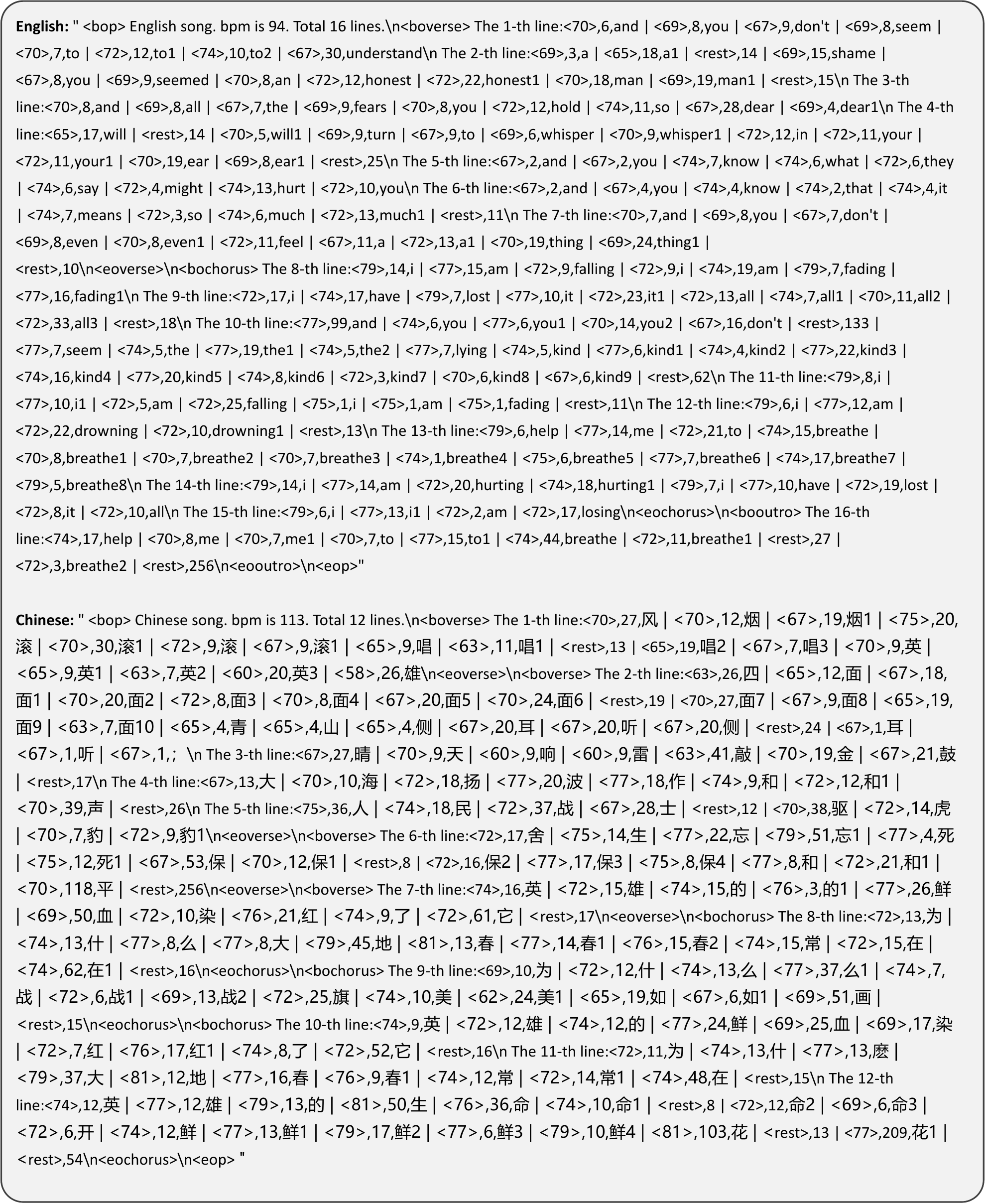}
	\caption{Two examples of lyric-melody pair data in English and Chinese with phrase-level tokens.}
	\label{fig:paired_case}
\end{figure*}

During the pretraining stage, we introduce three types of data. We give examples of what each format looks like.
As illustrated in Figure~\ref{fig:lyric_case}, we present Chinese and English instances of pure lyrics. The structure for pure melody is exemplified in Figure~\ref{fig:melody_case}. For lyric-melody pairs, bilingual versions are showcased in Figure~\ref{fig:paired_case}. 

\section{Baseline Construction}
\label{baseline_appendix}
\subsection{GPT}
\begin{figure*}[t!]
	\centering
	\includegraphics[width=0.9\textwidth]{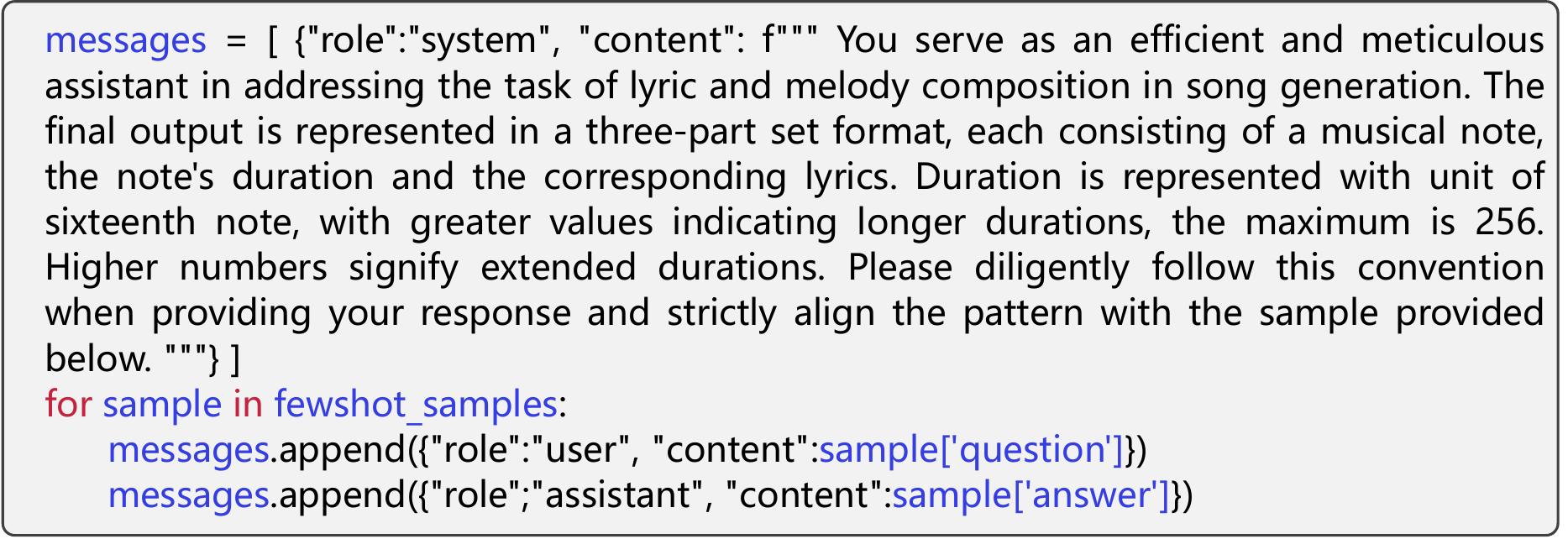}
	\caption{The prompt construction process for GPT-3.5/GPT-4, using few-shot in-context learning instructions.}
	\label{fig:gpt_prompt}
\end{figure*}

We invoke the GPT API to retrieve baseline results. We utilize a few-shot prompt to offer a template and instruct the model to follow suit. The pseudocode is illustrated in Figure~\ref{fig:gpt_prompt}.

\subsection{Open Source LLM}
For the open-source LLM, we select the base model as a fair comparison for all candidates. The prompt for the LLM is structured as follows:

\[
    \text{``system messages: } Q1 \rightarrow A1,  Q2 \rightarrow A2, \; Q3 \rightarrow \,"
\]

where system messages are the same as the one for GPT, \(Q1, Q2,\) and \(A1, A2\) are examples of the tasks we want the model to perform. We instruct the model to generate \(A3\) as the continuation of this prompt.



\section{Case Study: Evaluating Well-Structured Song Generation}
\label{sec:case_study}
\begin{figure*}[h]
	\centering
	\includegraphics[width=\textwidth]{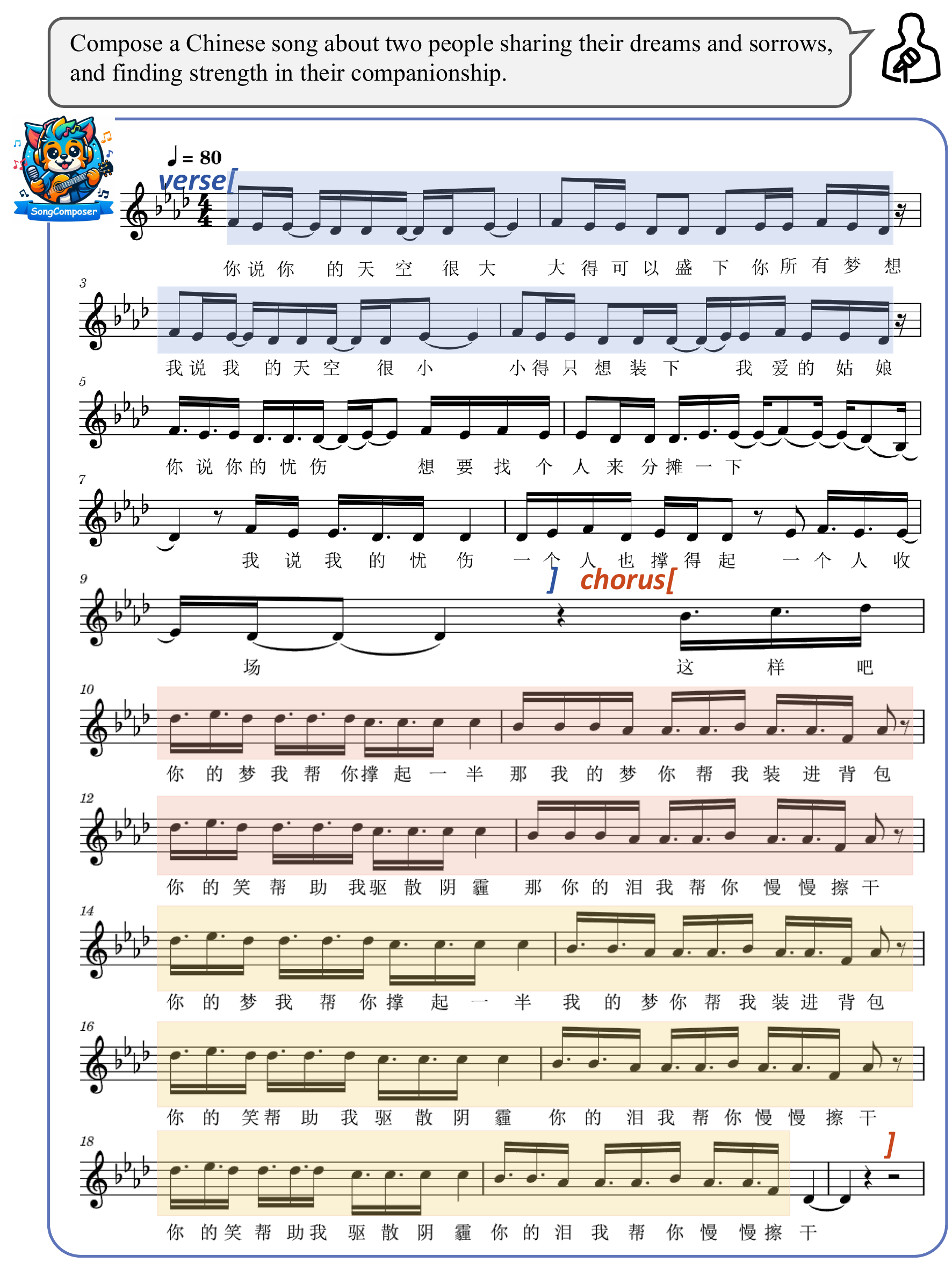}
	\caption{A text-to-song example in Chinese, featuring clear phrase-level repetitions highlighted with different colored boxes.}
	\label{fig:case_zh}
\end{figure*}

\begin{figure*}[h]
	\centering
	\includegraphics[width=\textwidth]{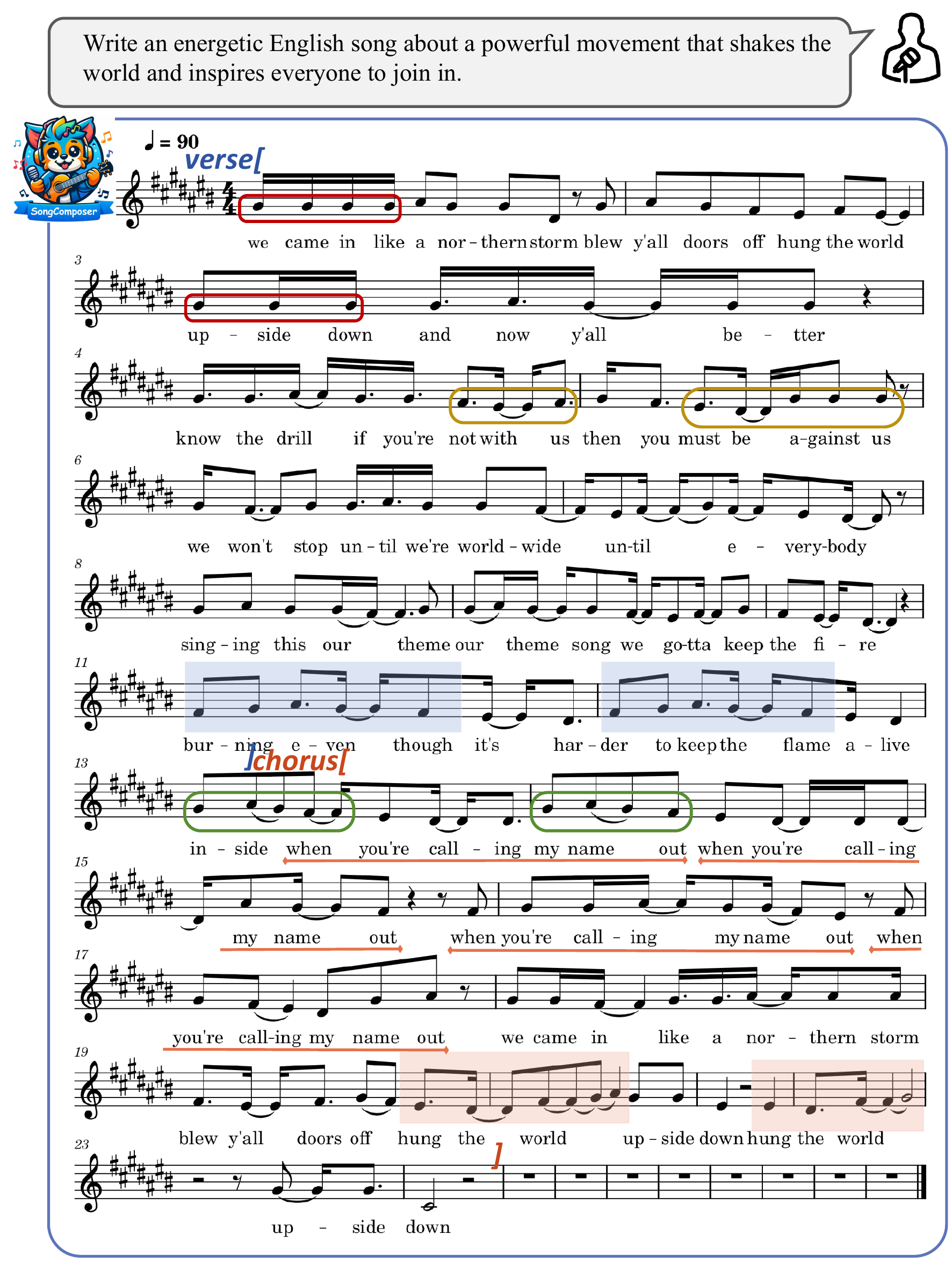}
	\caption{A text-to-song example in English, featuring prominent motif-level repetitions marked with different colored circles. }
	\label{fig:case_en}
\end{figure*}

To better validate SongComposer's ability to generate well-structured songs, we conducted a case study. Figures \ref{fig:case_zh} and \ref{fig:case_en} present examples of text-to-song generation in Chinese and English, respectively. We used different colored boxes to highlight phrase-level repetitions, different colored circles to mark motif-level repetitions, and underlines to indicate lyrical repetitions.

In these cases, we can observe distinct differences in SongComposer's handling of verses and choruses. Figure \ref{fig:case_zh} clearly exhibits phrase-level repetitions, while Figure \ref{fig:case_en} demonstrates significant motifs.
Notably, our lyrics harmonize with the melody, particularly in segments where the melody repeats, demonstrating semantic alignment.

\end{document}